# Study of the effect of external noise pickups on the performance of a cryogenic bolometer

A. Garai,[1,2] A. Reza,[3,a)] A. Mazumdar,[1,2] H. Krishnamoorthy,[1,2] G. Gupta,[3]
M.S. Pose,[3] S. Mallikarjunachary,[3] V. Nanal,[3] R. G. Pillay,[3,b)] and
S. Ramakrishnan[4]

[1]*India-based Neutrino Observatory, Tata Institute of Fundamental Research, Homi Bhabha Road, Mumbai 400005, India*
[2]*Homi Bhabha National Institute, V.N. Purav Marg, Anushaktinagar, Mumbai 400094, India*
[3]*Department of Nuclear and Atomic Physics, Tata Institute of Fundamental Research, Homi Bhabha Road, Mumbai 400005, India*
[4]*Deptment of Condensed Matter Physics & Materials Science, Tata Institute of Fundamental Research, Homi Bhabha Road, Mumbai 400005, India*

[a)]**Electronic mail**: ashifreza86@gmail.com
[b)]**Present address**: Department of Physics, Indian Institute of Technology Ropar, Rupnagar – 140001, India

This paper reports the detailed noise characterization, investigation of various noise sources and its mitigation to improve the performance of a cryogenic bolometer detector. The noise spectrum has been measured for a sapphire bolometer test setup with indigenously developed NTD Ge sensor in the CFDR system at Mumbai. The effect of external noise, arising either from ground loops in the system or from the diagnostic and control electronics of the cryostat, on the performance of a cryogenic bolometer is assessed. A systematic comparison of the influence of different noise pickups on the bolometer resolution is also presented. The best-achieved resolution ($\sigma_E$) at 15mK is ~15 keV for heater pulses and appears to be mainly limited by the noise due to the pulse tube cryocooler.

Over the last few decades, there is a strong interest in application of cryogenic bolometers in rare decay studies due to their excellent energy resolution, high sensitivity and a possibility to upscale the size[1-2]. Generally, the performance of cryogenic bolometers is not limited by the intrinsic resolution, but by the external noise sources in the system[3]. To achieve operating temperatures of 5-10mK, the pulsed tube based cryogenic systems, are attractive for long term operations in underground laboratories. In these systems, noise induced due to the mechanical vibrations is a major concern and attempts to reduce the same have been reported[4-5]. Other significant contributions arise from the mK thermometry (sensor+readout electronics) and system specific components like vacuum pumps and control units. In addition, minimization of ground loops is essential. With this motivation, we have studied the noise spectra of a test bolometer in the cryogen free dilution refrigerator.

The CFDR-1200 (Leiden Cryogenics) has been set up[6] at Tata Institute of Fundamental Research, Mumbai, for the prototype development of a Sn cryogenic bolometer (*TIN.TIN* -The INdia based TIN detector)[7]. To minimize the pulse tube vibration, a linear drive unit is used to smoothen the movement of motorized rotary valve control unit. The mixing chamber temperature ($T_{MC}$) is monitored by a Carbon Speer sensor, calibrated against a Cerium Magnesium Nitrate (CMN) thermometer. The CFDR setup is enclosed within a Faraday Cage to reduce the effect of EMI. The valve control unit of the pulse tube cooler is kept detached from the cryostat body to minimize the induced vibration. Presently, a test bolometer setup is developed comprising a sapphire absorber (20mm × 20mm × 0.4mm), indigenously fabricated NTD Ge sensor[8-9] and a heater element, which are strongly coupled to the absorber using a low-temperature Araldite. The absorber is weakly connected to the Cu block using 0.1mm thick Araldite dots (dia~1mm). The heater is developed by evaporating a 0.2μm thick Au meander on a Si substrate. The heater resistance, measured using an AC resistance bridge AVS-47B (PICOWATT), is ~ 0.6kΩ at 1K and remains constant down to 10mK. The NTD Ge sensors (6mm × 3mm × 1mm) are developed in-house. Measurements of the resistance of NTD Ge sensor ($R_S$) are done with AVS-47B for $R_S$<2MΩ. At lower temperatures, (i.e. $R_S$>2MΩ), $R_S$ is measured by applying a pseudo-constant current, generated by the voltage $V_B$ through a high resistor ($R_{B1}$=20GΩ), connected in series with the sensor (Fig. 1).

The resistance ($R_S$) for the NTD Ge Sensor (DB27) has been measured in the range of 10–400mK. It was observed that the measured resistance showed a deviation from the Mott behaviour[10] below 50mK and saturated at ~ 250MΩ, even though $T_{MC}$ cools down to 10mK. Therefore, noise spectra have been recorded to understand the effect of various external noise sources.

One of the major factors limiting the bolometer performance can be inefficient grounding leading to several ground loops. To overcome this, ground connections to various units are fanned out from a specially designated clean earth pit. The chassis acts as a master ground for the DAQ box, the amplifier, the AVS preamplifier as well as the cryostat (see Figure 2). The CFDR cryostat, the differential amplifier and DAQ signal box, AVS preamplifiers for diagnostic thermometry and the sensor readout are enclosed within the Faraday Cage. The CFDR controls for $^3$He-$^4$He gas handling system and accessories are routed through an optically isolated USB-RS232 interface. The NI-PXI chassis, AVS bridges and PCs are fitted in an anodized metallic rack. The PXI is connected to the PC with an optical link to reduce the EMI pickups.

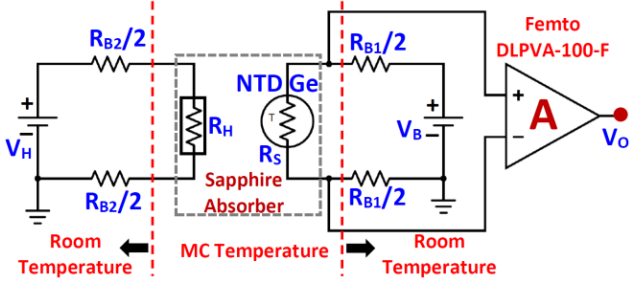

**FIG. 1.** Schematic circuit for the sapphire bolometer readout.

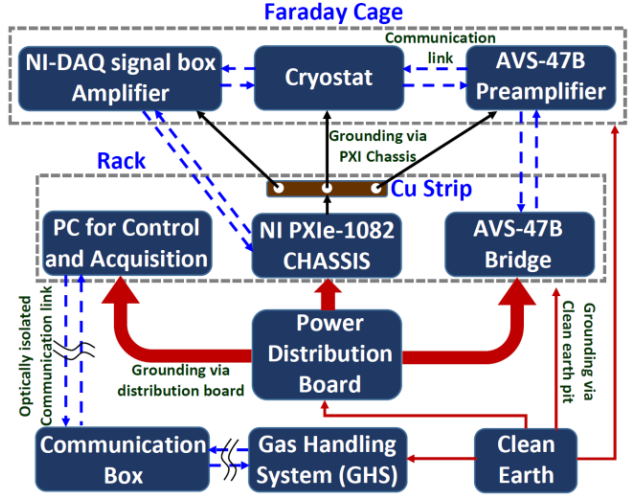

**FIG. 2.** A schematic layout of optimized ground configuration.

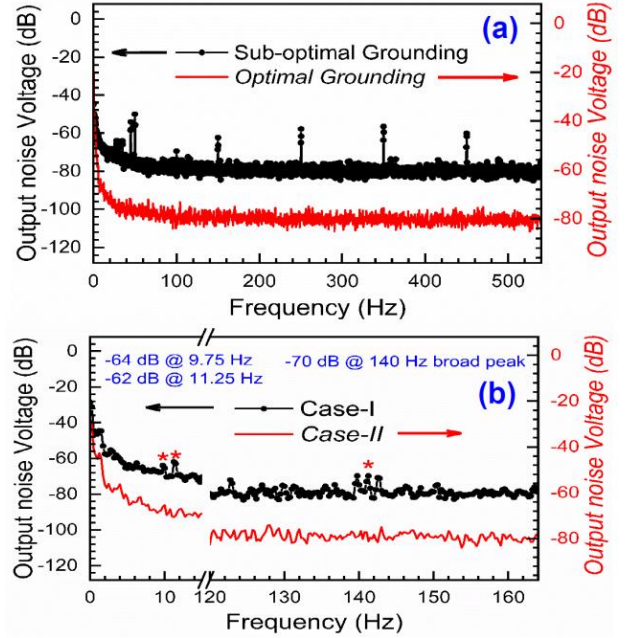

**FIG. 3.** FFT spectra of DB27 output showing (a) Effect of ground loops on 50Hz noise at $T_{MC}$ = 15 mK (b) Effect of pulse tube induced noise.

As an example of a ground loop in the sub-optimal grounding, the master ground is assigned to the cryostat, which results in multiple ground connections for DAQ– through the cryostat and through the power connection. In this configuration, 50Hz and its harmonics are clearly visible in the noise spectra (Fig. 3a). As a result, the $T_{MC}$ could not be stabilized below 15mK.

Further noise measurements have been carried out with optimal grounding at $T_{MC}$=10mK for four configurations: *Case-I*: pulse tube in normal drive, motor head mounted on the cryostat; *Case-II*: pulse tube in linear drive, motor head detached from the cryostat; *Case-III*: Case-II configuration with gauge electronics of Inner Vacuum Chamber disconnected; *Case-IV*: Case-II configuration with electronics of all vacuum gauges disconnected. In all four cases, noise spectra are measured over a frequency range of 0–25kHz with a voltage gain of 80dB. Noise peaks at 9.75Hz, 11.25Hz and 140Hz seen in Fig. 3b for Case-I are mainly contributed by the harmonics of pulse tube induced noise of 1.4Hz, which disappears for linear drive mode in Case-II. This helps in cooling the sensor, which is also reflected in the measured resistance of 370MΩ as compared to 250MΩ for Case-I. However, no appreciable difference in the FFT spectra above 500Hz is observed between Case-I and Case-II. Vacuum pumps and gauges are found to contribute to high frequency pickups at 827Hz, 985.5Hz and 16kHz. Details of the various noise pickups are summarized in Table I. In Case-IV, the measured resistance is 594MΩ.

Figure 4(a) shows that the sensor resistance has increased substantially in Case-IV as compared to Case-I and the sensor follows Mott behavior down to 40mK indicating a net improvement in reducing the heat leak to the sensor. The effect of ground loops was also reflected in the minimum $T_{MC}$ and in the measured resolution. For Case-IV, $T_{MC}$ ~ 5mK could be achieved as compared to 6.7mK in the sub-optimal case.

TABLE. 1. Output voltage in dB (text in italics) for different frequencies as measured from the noise spectra. (BBN: Below Baseline Noise)

| Different Cases | Frequency (Hz) | | | | |
|---|---|---|---|---|---|
| | 9.75/11.25/140 | 543.5 | 827.5 | 985.5 | 15.73k/16.27k |
| Case-I | -64/-62/-70 | -63 | -64 | -62 | -55 |
| Case-II | BBN | -62.5 | -65.7 | -58.3 | -59 |
| Case-III | BBN | -62 | -70.4 | -60.4 | BBN |
| Case-IV | BBN | -63 | -71.6 | -60.2 | BBN |

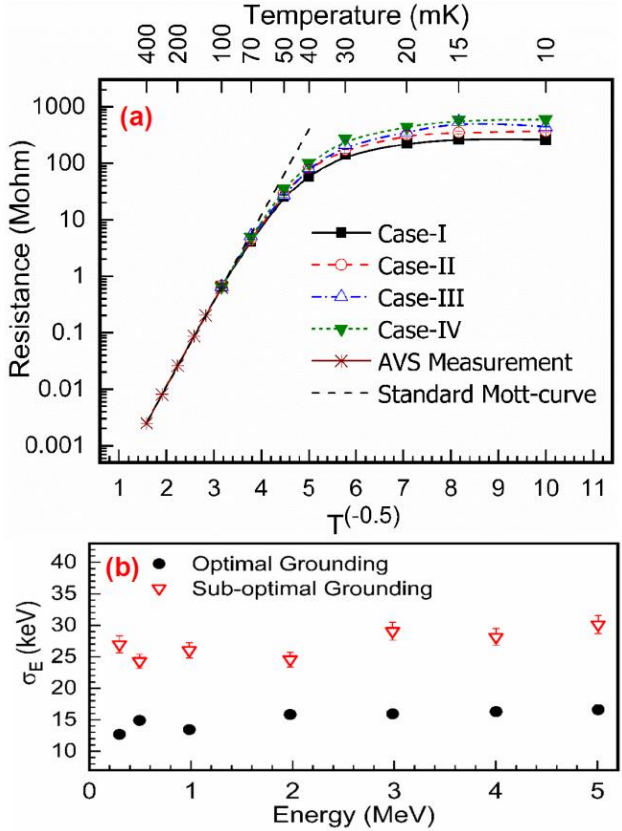

**FIG. 4.** (a) Measured resistance of DB27 for T=10 – 400 mK together with standard Mott curve fit ($R_0$ = 10.2 (0.5) $\Omega$ and $T_0$ = 12.2 (0.2) K) (b) Measured bolometer resolution with the phonon signal.

To measure the bolometer resolution, an external phonon signal is generated, by connecting a 2M$\Omega$ resistor ($R_{B2}$) in series with the heater (Fig. 1). The pulses of varying amplitude, 200μs wide with a repetition rate of 2Hz, are supplied to the heater to simulate phonon signal. The voltage signal from the sensor is analyzed off-line[11]. Figure 4(b) shows that the measured resolution at 15mK for Case-IV, which worsens by ~80% in case of the sub-optimal grounding. The resolution is found to improve by ~30% when the pulse tube is switched from normal drive to linear drive. However, different configurations (Case-II to Case-IV) yield similar results (~15keV) within measurement errors.

In summary, we have presented the influence of various noise sources on the performance of a sapphire bolometer in the CFDR system inclusive of control, diagnostic thermometry and DAQ. It is shown that the presence of ground loops can worsen the bolometer resolution ~80%. Further, the noise pickups from vacuum pumps and gauges also introduce thermal load on the sensor. In the best configuration, $\sigma_E$~15keV is obtained for the sapphire bolometer at 15mK.

The authors would like to thank *TIN.TIN* and INO collaboration for their support and Mr. K.V. Divekar for assistance during measurements.

[1]P. F. de Salas et al., Front. Astron. Space Sci. **5**, 1 (2018).
[2]C. Alduino et al., Phys. Rev. Lett. **120**, 132501 (2018).
[3]C. Enss et al., J. Low Temp. Phys. **151**, 5 (2008).
[4]A. M. J. den Haan et al., Rev. Sci. Instrum. **85**, 035112 (2014).
[5]A. D'Addabbo et al., Cryogenics **93**, 56 (2018).
[6]V. Singh et al., Pramana **81**, 719 (2013).
[7]V. Nanal, EPJ Web Conf. **66**, 08005 (2014).
[8]S. Mathimalar et al., 11th International Workshop on Low Temperature Electronics, July, 7 - 9, France (2014).
[9]A. Garai et al., J. Low Temp. Phys. **184**, 609 (2016).
[10]E. Pasca et al., Proceedings of 8th International Conference on Advanced Technology and Particle Physics, **2**, 93 (2004).
[11]A. Garai et al., Proceedings of DAE International Symposium on Nuclear Physics, **63**, 1140 (2018).